\documentclass{aastex61}

\received{mm dd, yyyy}
\revised{mm dd, yyyy}
\accepted{mm dd, yyyy}
\submitjournal{ApJ}

\shorttitle{Coronal jet and subsequent two-ribbon flare}
\shortauthors{Joshi et al.}

\begin{document}
\title{Observational and model analysis of a two-ribbon flare possibly induced by a neighbouring blowout jet}

\correspondingauthor{Bhuwan Joshi}
\email{bhuwan@prl.res.in}

\author{Bhuwan Joshi}
\affil{Udaipur Solar Observatory, Physical Research Laboratory, Udaipur 313 001, India}

\author{Julia K. Thalmann}
\affil{Institute of Physics, University of Graz, Universit$\ddot{a}$tsplatz 5, A-8010 Graz, Austria}

\author{Prabir K. Mitra}
\affil{Udaipur Solar Observatory, Physical Research Laboratory, Udaipur 313 001, India}

\author{Ramesh Chandra}
\affil{Department of Physics, DSB Campus, Kumaun University, Nainital 263 002, India}

\author{Astrid M. Veronig}
\affil{Institute of Physics, University of Graz, Universit$\ddot{a}$tsplatz 5, A-8010 Graz, Austria}

\begin{abstract}

In this paper, we present unique observations of a blowout coronal jet that possibly triggered a two-ribbon confined C1.2 flare in a bipolar solar active region NOAA 12615 on 2016 December 5. The jet activity initiates at chromospheric/transition-region heights with a small brightening that eventually grows in a larger volume with well developed standard morphological jet features, viz., base and spire. The spire widens up with a collimated eruption of cool and hot plasma components, observed in the 304 and 94~\AA~channels of AIA, respectively. The speed of the plasma ejection, which forms the jet's spire, was higher for the hot component ($\sim$200~km~s$^{-1}$) than the cooler one ($\sim$130~km~s$^{-1}$). 
The NLFF model of coronal fields at pre- and post-jet phases successfully reveal opening of previously closed magnetic field lines with a rather
inclined/low-lying jet structure.
The peak phase of the jet emission is followed by the development of a two-ribbon flare that shows coronal loop emission in HXRs up to $\sim$25 keV energy. The coronal magnetic fields rooted at the location of EUV flare ribbons, derived from the NLFF model, demonstrate the pre-flare phase to exhibit an ``X-type" configuration while the magnetic fields at the post-flare phase are more or less parallel oriented. The comparisons of multi-wavelength measurements with the magnetic field extrapolations suggest that the jet activity likely triggered the two-ribbon flare by perturbing the field in the interior of the active region.

\end{abstract}


\keywords{Sun: activity -- Sun: Corona -- Sun: flare -- Sun: X-rays, gamma-rays}

\section{Introduction} \label{sec:intro}
Transient eruptions of plasma are observed at a wide range of spatial, temporal, and energy domains in the solar atmosphere . The scale of plasma ejections varies from a huge coronal mass ejections (CME) to small-scale phenomena of coronal jets. 
Although the energy involved in coronal jets is smaller than that of nominal solar flares and CMEs, they are thought to play an important role as a source of mass and energy input to the upper solar atmosphere \citep[see review by][]{Raouafi2016}. The multi-wavelength observations of coronal jets provide opportunities to explore the origin of small-scale transients in the solar atmosphere and their relation to the underlying magnetic reconnection process. 

The extensive investigations of coronal jets began with the observations from the Soft X-ray Telescope (SXT) on board Yohkoh \citep{Tsuenta1991}. The soft X-ray (SXR) images revealed several jet-like features, i.e., transient, dynamic, and collimated structures in the solar corona showing enhanced X-ray emission \citep{Shibata1992}. Using Yohkoh observations, \cite{Shimojo1996} studied a large data set of solar X-ray jets that include events from different regions of the solar disk (i.e., active region, quiet region and coronal hole). They found jets to have lengths of few $\times$ 10$^{4}$--4 $\times$10$^{5}$~km with an average value of 1.5 $\times$ 10$^{5}$~km; widths 5 $\times$ 10$^{3}$--10$^{5}$~km with an average value of 1.7 $\times$ 10$^{4}$ km; velocities 10--1000~km~ s$^{-1}$ with an average value of $\sim$ 200~km~s$^{-1}$; lifetimes from few minutes to several hours. Based on the physical and morphological properties of X-ray jets along with the magnetic characteristics of jet producing active regions \citep{Shimojo1998}, a ``standard jet" model was proposed which is based on magnetic reconnection. In the standard jet model, the magnetic reconnection between a newly emerging bipole and pre-existing locally unipolar magnetic field region leads to jet activity \citep[][]{Yokoyama1995,Shimojo1996,Canfield1996,Shimojo2000}. 

After the discovery of solar X-ray jets from Yohkoh, the jet activity was systematically examined at various energy and wavelength channels. \cite{Wang1998} and \cite{Wang2002} studied jet phenomena in the extended coronal layers ($>$ 2 R$_{\odot}$). They identified the source of white light coronal jets using data from the Extreme-ultraviolet Imaging Telescope (EIT) and the Large Angle Spectrometric Coronagraph (LASCO) on board Solar and Heliospheric Observatory (SOHO) and confirmed the white light jets to be the outward extensions of EUV jets. By combining high resolution EUV images from the Transition Region and Coronal Explorer (TRACE) with Yohkoh/SXT measurements, \cite{Alexander1999} found collimated ejection of both hot and cold material during the jet activity. \cite{Jiang2007} presented a detailed investigation of a coronal jet using EUV (from TRACE), soft X-ray (from Yohkoh), and H$\alpha$ data, and demonstrated that the evolution of hot and cool components of the jet differ spatially and temporally. The coronal jets have also been widely studied by analyzing data from the Solar Terrestrial Relations Observatory (STEREO). \cite{Patsourakos2008} detected helical structure in a polar coronal jet from two viewpoints observations of the Extreme Ultraviolet Imaging Telescope (EUVI) on board the twin STEREO spacecrafts (i.e., STEREO-A and -B). \cite{Nistico2009, Nistico2010} produced a catalogue of coronal jets that appeared in polar and equatorial coronal hole regions, respectively, using data from STEREO satellites. Their studies suggest no substantial physical differences between coronal jets at polar and equatorial coronal holes. By examining many X-ray jets in Hinode/X-ray telescope (XRT) coronal X-ray movies, \cite{Moore2010} introduced a new variant of X-ray jets that exhibit much broader spire with enhanced emission from the extended bright region at its base. This new class of jets was classified as ``blowout jets". Observations from the 304~\AA\ channel of the Atmospheric Imaging Assembly (AIA) on the Solar Dynamics Observatory (SDO), which images plasma at cooler transition region temperatures (T~$\sim$~10$^{5}$~K) within the jet structure, further confirmed the dichotomy of standard and blowout jets in regard to their morphology and dynamics \citep{Moore2013}. Observations also indicate role of blowout jet toward triggering of large active region filament eruption \citep{Joshi2016}. Sometimes, active regions exhibit repeated jet activity, spanning over several hours, which are observed to be associated with localized episodes of flux emergence or cancellation \citep{Chandra2015,JoshiR2017}. 
 
Different forms of transient solar activity - e.g., jets, flares, and coronal mass ejections - involve energy release from localized regions of the solar atmosphere at very different temporal, spatial and spectral scales \citep[see e.g., ][]{Shibata1999}. 
Magnetic reconnection is a generally accepted mechanism to explain the energy release in these events. However, the significant changes in the morphology, dynamics, and energetics of different kinds of solar transients point toward the diversities and complexities of the underlying physical processes which eventually depends on the magnetic field configurations of the corona, and various parameters of magnetic field and plasma from the sub-photospheric layers to the outer corona \citep[for a review, see][]{Wiegelmann2014}. During a solar flare, the coronal magnetic reconnection gives rise to bright emissions from region close to the apex of coronal loops (in the form of X-ray looptop sources) as well as their feet (in the form of chromospheric flare ribbons or hard X-ray foot-point sources) \citep[see review by][]{Benz2017}. On the other hand, coronal jets occur at much smaller energy and spatial scales. Although the reconnection is a common mechanism for energy release and plasma flows in jet and flare/CME models, our understanding is still incomplete about the triggering of reconnection at the first place and its role in driving various eruptive activities.
Here the question arises: whether the basic nature of solar transients - irrespective of their length or time scales - is similar or not? 
To answer these questions we need thorough investigations of suitable multi-wavelength observations and magnetic field modeling. Now, with
space-bourne multi-wavelength imaging observations of the Sun at high spatial and time cadence, it has become possible to meticulously examine the cause and consequences of the transient energy release and to investigate similarities between various reconnection-driven phenomena in the solar atmosphere.  

In this paper, we analyze a coronal jet that possibly triggers a two-ribbon flare. The activities occurred in a simple bipolar solar active region NOAA 12615 on 2016 December 5. 
The event occurred during the decline phase of the solar cycle 24 when the activity was very low. Notably, NOAA 12615 produced only C and B class flares. Due to the low level of activity from NOAA 12615, the background emissions from the active region were very low at the time of flare onset. This has enabled us to examine features of the early eruption as well as flare emissions with high sensitivity. The precise estimations of the temporal evolution of various eruptive phenomena, mainly during the pre-flare to early impulsive phases of the flare, provide insight into the triggering mechanism of the eruption along with the underlying magnetic reconnection process. Although the jet eruption and resulting flare studied here were of smaller spatial and temporal scales, such observations can also help in understanding the physics of large flux rope eruptions. 
Section~\ref{sec:data} provides a brief record of data sources and analysis methods used in this study. In section~\ref{sec:analysis}, we present a detailed description of multi-wavelength data set and derive observational results. The analysis of magnetic field modeling of the active region is described in Section~\ref{sec:mag_model}. Interpretations on the jet and flare activity along with their possible interrelations are discussed in Section~\ref{sec:discussion}. 
Major highlights of the study are given in
Section~\ref{sec:conclusion}.

\section{Data and methods}
\label{sec:data}

This paper is preliminary based on the observations from the Atmospheric Imaging Assembly \citep[AIA;][]{Lemen2012} on board the Solar Dynamics Observatory \citep[SDO;][]{Pesnell2012}.  AIA observes the full solar disk at 12-s cadence in seven EUV filters (94, 131, 171, 193, 211, 304, and 335~\AA), at 24-s cadence in two UV filters (1600 and 1700~\AA), and at 3600-s cadence in the white light filter (4500~\AA). In this paper, we present AIA images taken in the 94, 171, 304, and 1600~\AA~band passes with a spatial resolution of 0$\arcsec $.6 pixel$ ^{-1} $. 

The Reuven Ramaty High Energy Solar Spectroscopic Imager \citep[RHESSI;][]{LinRP2002} provided uninterrupted observations of the Sun during the jet and flare activty. RHESSI observes the full Sun with an unprecedented combination of spatial resolution (as fine as $ \sim $2$ {\arcsec} $.3) and energy resolution (1--5~keV) in the energy range 3~keV to 17~MeV.
It is noteworthy that RHESSI's attenuator state was A0 during the whole observing period, i.e., RHESSI observed with its highest sensitivity at low energies. This has enabled us to examine the X-ray emission associated with the event down to 3~keV energy. The image reconstruction is done with CLEAN algorithm using front detector segments only. Out of 9 detector segments, segment 2 was excluded for imaging at 6--12 and 12--25 keV while segments 2 and 7 were excluded for 3--6 keV imaging.

To study the photospheric structure of the active region, we employ magnetogram and intensity images from Helioseismic and Magnetic Imager \citep[HMI;][]{Schou2012} on board SDO. HMI images have spatial resolution of 0$\arcsec $.5 pixel$ ^{-1}$.

We use  full-disk  HMI {\sf hmi.B\_720s} observables  \citep{2014SoPh..289.3483H} to model the 3D coronal magnetic field configuration (total field strength, inclination and the 180$^\circ$ ambiguity corrected azimuth) to derive the image-plane magnetic field vector field with a native (full-resolution) plate scale of  $\approx$1\farcs.
The latter is projected to a local reference frame following \cite{1990SoPh..126...21G}. We bin the magnetic field data to a lower resolution ($\approx$2\farcs~pixel$^{-1}$) and use a  sub-field, covering the flaring AR as well as its quiet-Sun surrounding, as  input to a nonlinear force-free (NLFF) model scheme (for details see  \cite{2010A&A...516A.107W} and Sect.\ 2.2.1 of \cite{2015ApJ...811..107D}). 

\section{Analysis and results}
\label{sec:analysis}

\subsection{Activity of NOAA 12615: Morphology and spatial evolution}
\label{sec:EUV_imaging}

\begin{figure}[ht!]
\plotone{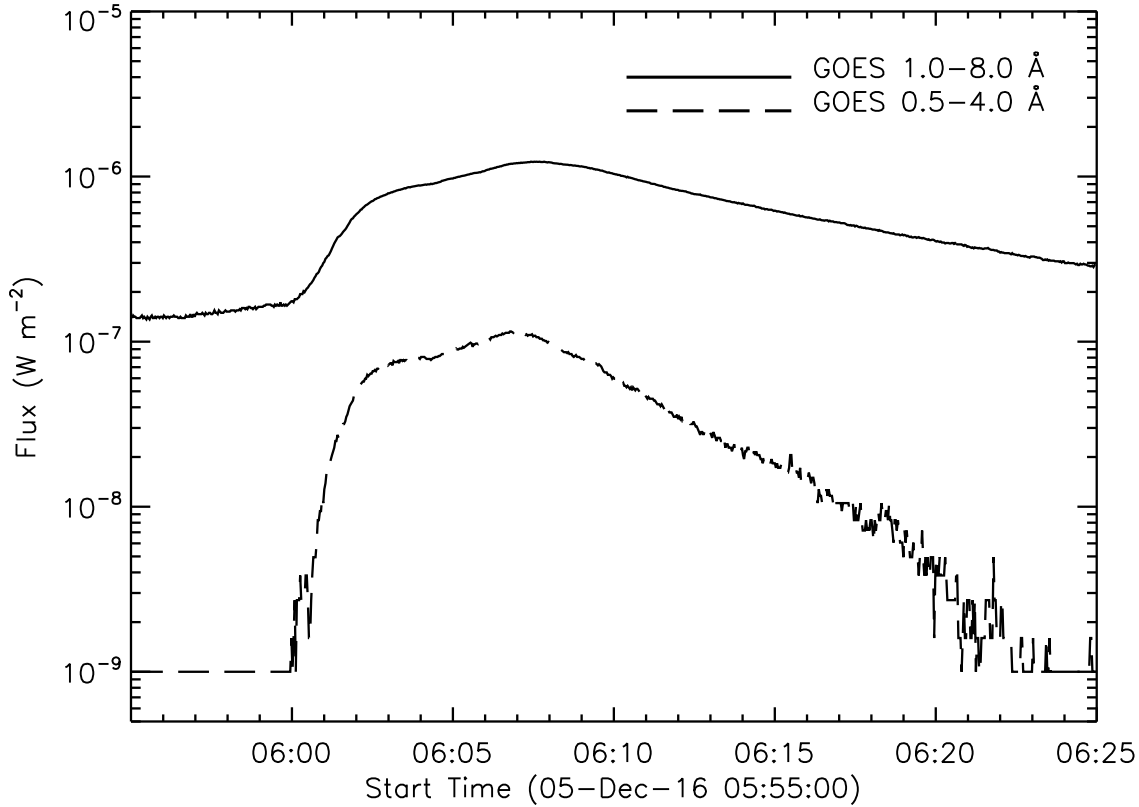}
\caption{GOES time profiles showing the evolution of solar X-ray flux in 0.5--4 and 1--8~\AA~energy channels. \label{fig:goes_lc}}
\end{figure}

\begin{figure}[ht!]
\epsscale{0.55}
\plotone{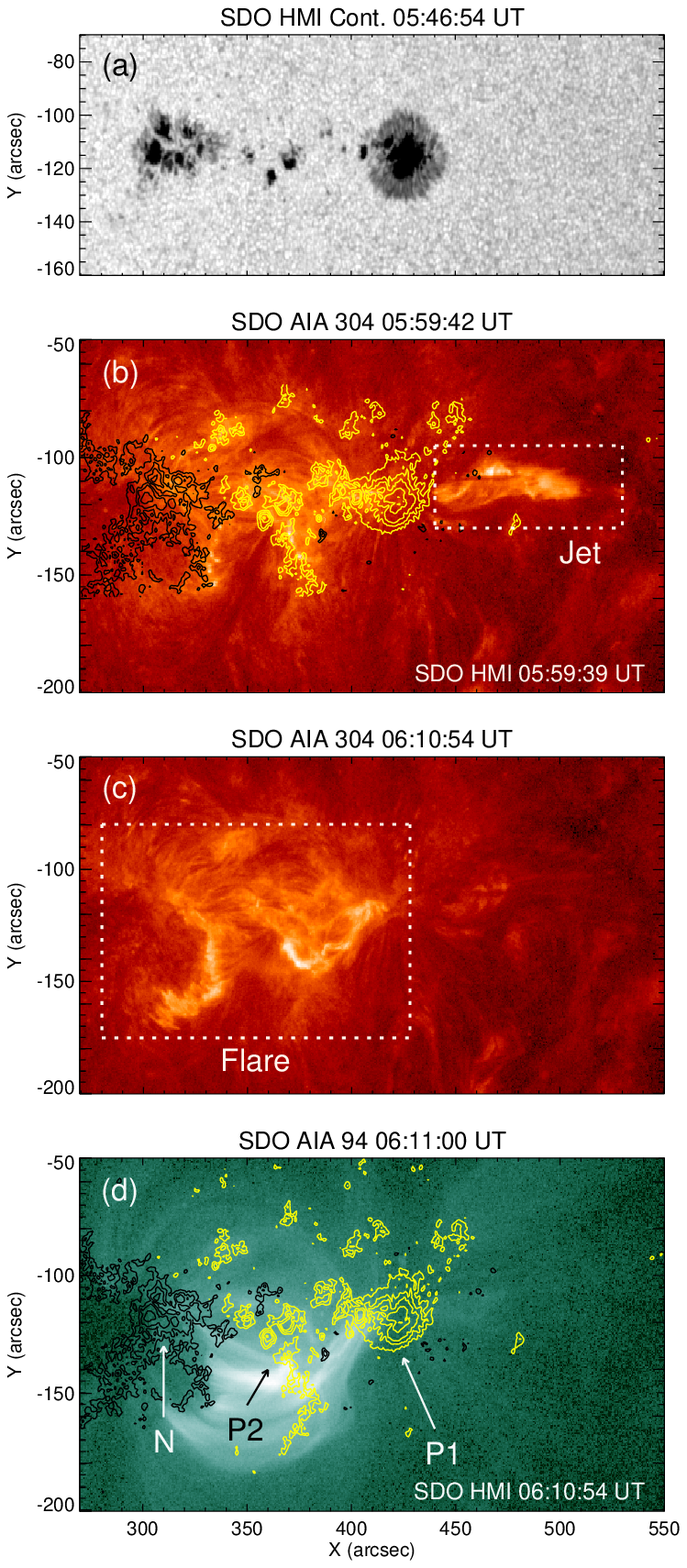}
\caption{Panels (a): HMI intensity image of the active region showing the leading and following sunspot groups. Panels (b) and (c): AIA 304~\AA~images showing jet and flare processes. Dotted rectangular boxes mark jet and flare regions. Panel (d): AIA 94~\AA~image showing the post-flare loop configuration. In panels (b) and (d), AIA images are overplotted by HMI Line-of-sight (LOS) magnetogram to understand the distribution of photospheric magnetic field with respect to the flare signatures at corona and chromosphere-transition region heights.
Yellow and black contours represent positive and negative magnetic polarities, respectively. Time indicated at the bottom right corner in panels (b) and (d) is the time of SDO/HMI map. The HMI contour levels are $\pm$200, $\pm$600, $\pm$1000, $\pm$ 1400 G. The prominent photospheric magnetic flux domains involved in the post-flare loop system are marked as P1, P2, and N in panel (d).  
\label{fig:AR_mag_euv}}
\end{figure}

We have thoroughly analyzed the morphological and kinematical characteristics of the coronal jet and associated C1.2 flare using AIA multi-channel observations. We mainly emphasize the observations in the 304~and 94~\AA\ bands for a detailed understanding of the cool and hot plasma components, respectively, associated with transient activities. The AIA 304~\AA~channel (He~{\footnotesize II}; log(T)=4.7) images the solar structures formed at the chromosphere and the transition region. Observations from the AIA 94~\AA~channel (Fe~{\footnotesize XVIII}; log(T)=6.8) provide diagnostics of high temperature plasma structures associated with the flaring corona.

\begin{figure}[ht!]
\plotone{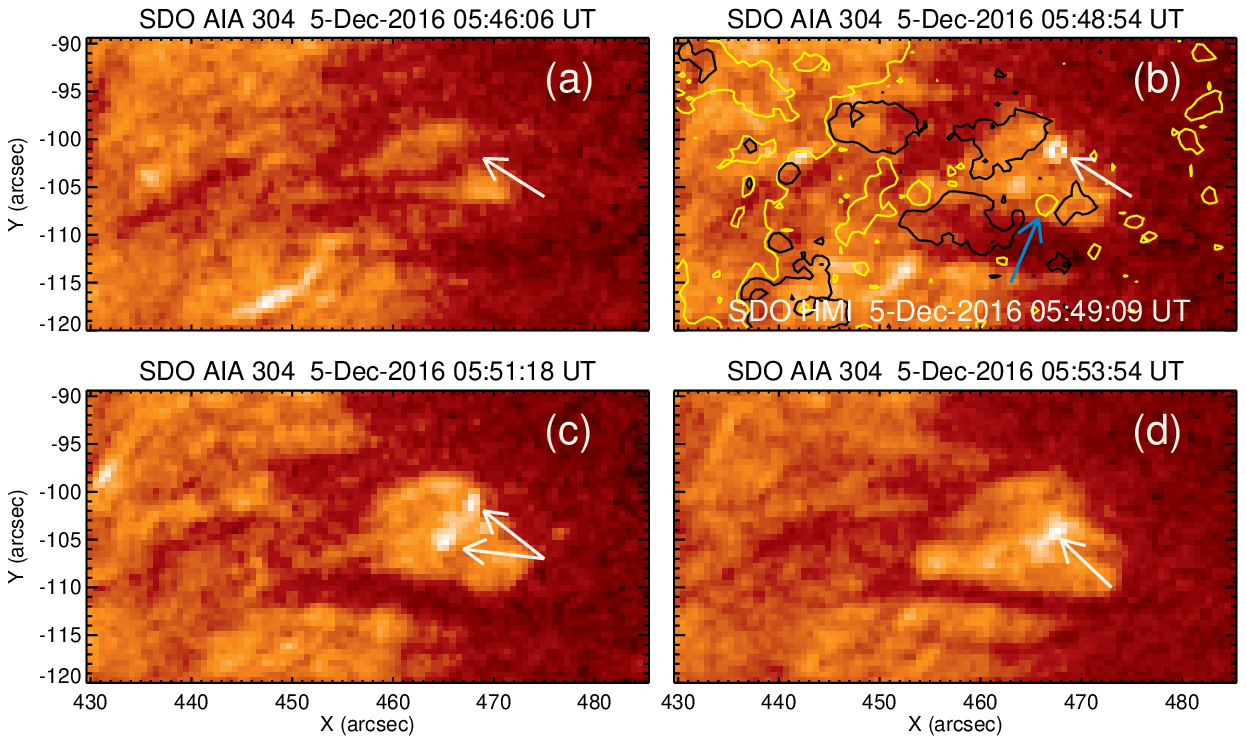}
\caption{AIA 304~\AA~images showing the initiation phase of the jet. The earliest brightening, forming the jet's base, started as early as $\sim$05:48~UT (cf. region marked by white arrow in panels (a) and (b)). Yellow and black contours in panel (b) represent positive and negative magnetic polarities, respectively. The contour levels are $\pm$25 and $\pm$500~G. Time of SDO/HMI map is indicated at the bottom of the panel. Evolution of the jet is shown by white arrows in panels (b)--(d). Blue arrow is panel (b) indicates a patch of positive polarity magnetic flux (yellow contour) which disappears during the jet activity.
\label{fig:pre-jet_AIA_HMI}}
\end{figure}

In Figure~\ref{fig:AR_mag_euv}, we show the main features associated with the jet and flare activity with respect to the photospheric structure of the active region. The jet activity precedes a two-ribbon flare. The jet is located to the west of the flaring region (Figure~\ref{fig:AR_mag_euv}(b)). Based on the morphology of flare ribbons (Figure~\ref{fig:AR_mag_euv}(c)) and post-flare loops (Figure~\ref{fig:AR_mag_euv}(d)), we identify the magnetic connectivity between a negative polarity region (N) with two distinct positive polarity regions (P1 and P2; See Figure~\ref{fig:AR_mag_euv}(d)). We start with a description of the jet-associated low-atmospheric emission. In Figure~\ref{fig:pre-jet_AIA_HMI}, we visualize the early stages of the jet evolution, as observed at chromospheric temperatures. First jet-related brightenings, in the form of a small-scale loop-like structure, were observed as early as 05:48~UT (positioned at around $x$=468\farcs, $y$=-101\farcs; Indicated by white arrow in Figure~\ref{fig:pre-jet_AIA_HMI}(b)). Comparison to the underlying line-of-sight magnetic field (black and yellow contours) shows that these early brightening is located in a region of small-scale magnetic flux of mixed polarity. We note that the location of the initial brightening coincided with a region of disappearing positive line-of-sight magnetic flux (indicated by a blue arrow in Figure~\ref{fig:pre-jet_AIA_HMI}(b)).

\begin{figure}[ht!]
\plotone{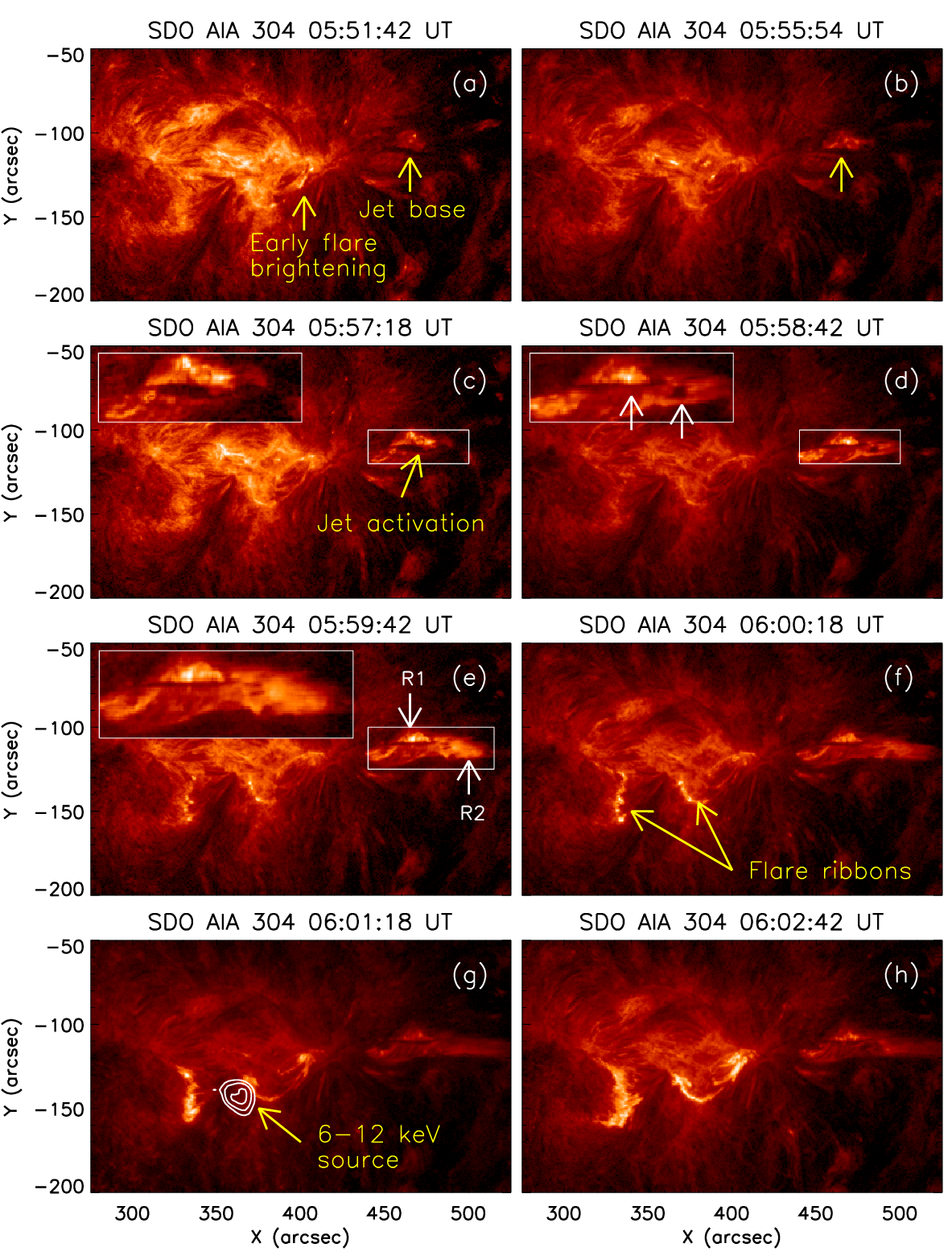}
\caption{Sequence of AIA~304~\AA~images showing the jet activity and subsequent flare developments.  In panel (g), we show the spatial location of flaring X-ray emission with respect to EUV~304~\AA~structures by overplotting co-temporal RHESSI 6--12 keV (red) contours on AIA image. The contour levels are set as 40\%, 60\%, and 90\% of the peak flux. The brightenings at two distinct regions of the blowout jet are marked by R1 and R1 in panel (e). The inlet at the top left corner of panels (c), (d), and (e) show zoomed images of a selected region (shown by smaller rectangular area) from the corresponding panel.
\\
(An animation of this figure is available (contact: bhuwan@prl.res.in))
\label{fig:aia304}}
\end{figure}

In Figure~\ref{fig:aia304}, we present a sequence of AIA 304~\AA~images depicting both, the jet and subsequent flare activity. The initial loop-like brightening spreads after $\sim$05:54~UT and forms the base of the jet (Figure~\ref{fig:aia304}(b)). The jet region grows rapidly in size and intensity. We also notice a small filament-like dark feature at this location (see the jet activation region in Figure~\ref{fig:aia304}(c)). From $\sim$05:57~UT onward, we observe fast and nearly collimated ejection of bright as well as dark material which forms the spire of the jet (Figure~\ref{fig:aia304}(c)-(h)). As the plasma eruption continues through the jet's spire, two bright flare ribbons form to the east of the sunspot after
$\sim$05:59~UT (Figure~\ref{fig:aia304}(e)-(h)). It is noteworthy that 
enhanced emission in the form of flare kernels can be observed in that area already at $\sim$05:50~UT, i.e., co-temporal with the jet activation
(see location marked as early flare brightening in Figure~\ref{fig:aia304}(a)). Such features intermittently occurred also at other locations within the flaring area and before the development of prominent flare associated structures, such as, flare ribbons, hot coronal loops, etc.. At the time when the jet-related enhanced emission is ceasing (after $\sim$06:02~UT), strong emission still stems from the location of the flare ribbons.

\begin{figure}[ht!]
\epsscale{.75}
\plotone{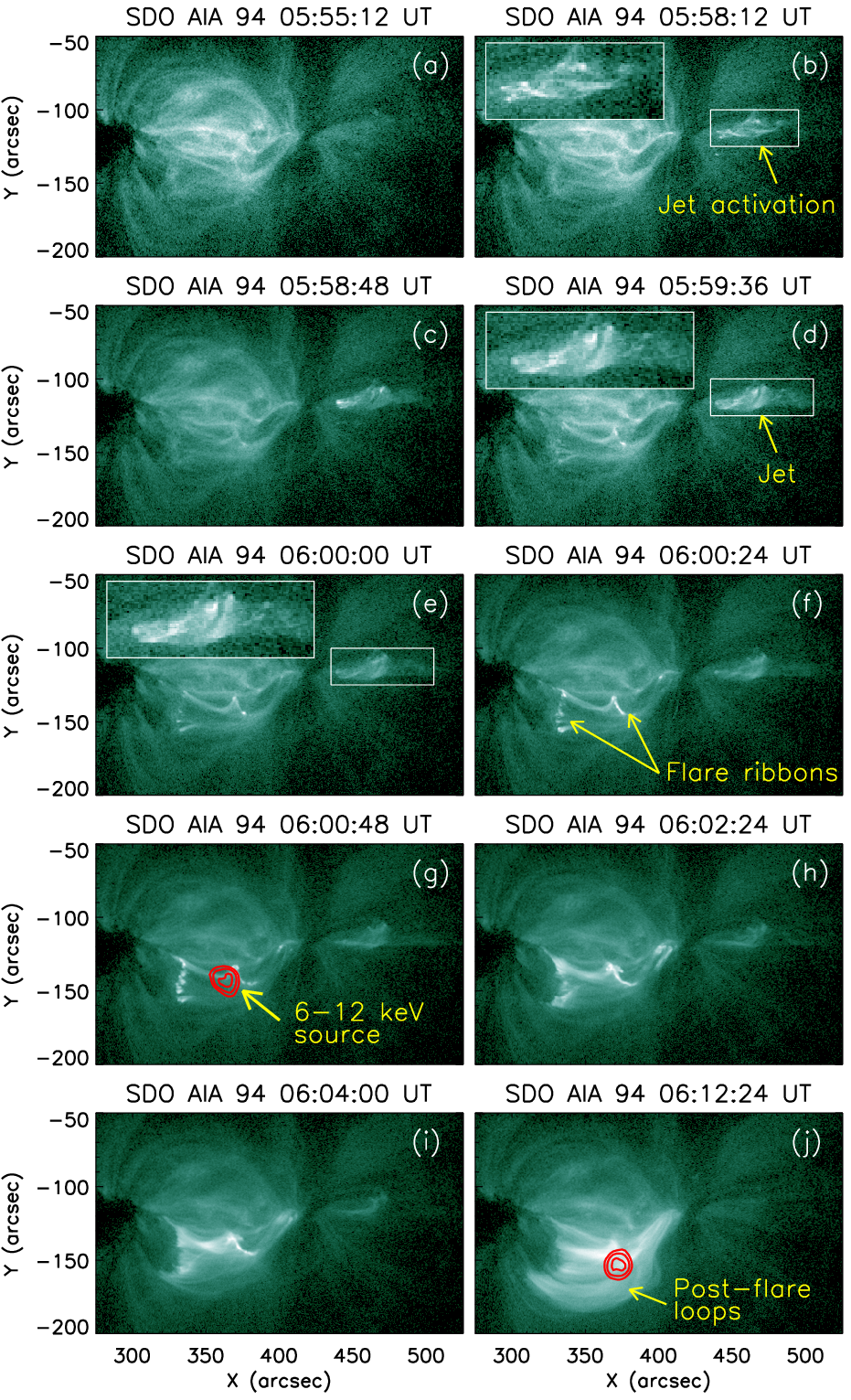}
\caption{Sequence of AIA~94~\AA~images showing evolutionary phases of the coronal jet and subsequent flare emission. To show the spatial location of flaring X-ray emission with respect to EUV~94~\AA~structures, we overplot co-temporal RHESSI 6--12 keV (red) contours on AIA image in panels (g) and (j). The contour levels are set as 40\%, 60\%, and 90\% of the peak flux. The inlet at the top left corner of panels (b), (d), and (e) show zoomed images of a selected region (shown by smaller rectangular area) from the corresponding panel.
\label{fig:aia94}}
\end{figure}

\begin{figure}[ht!]
\plotone{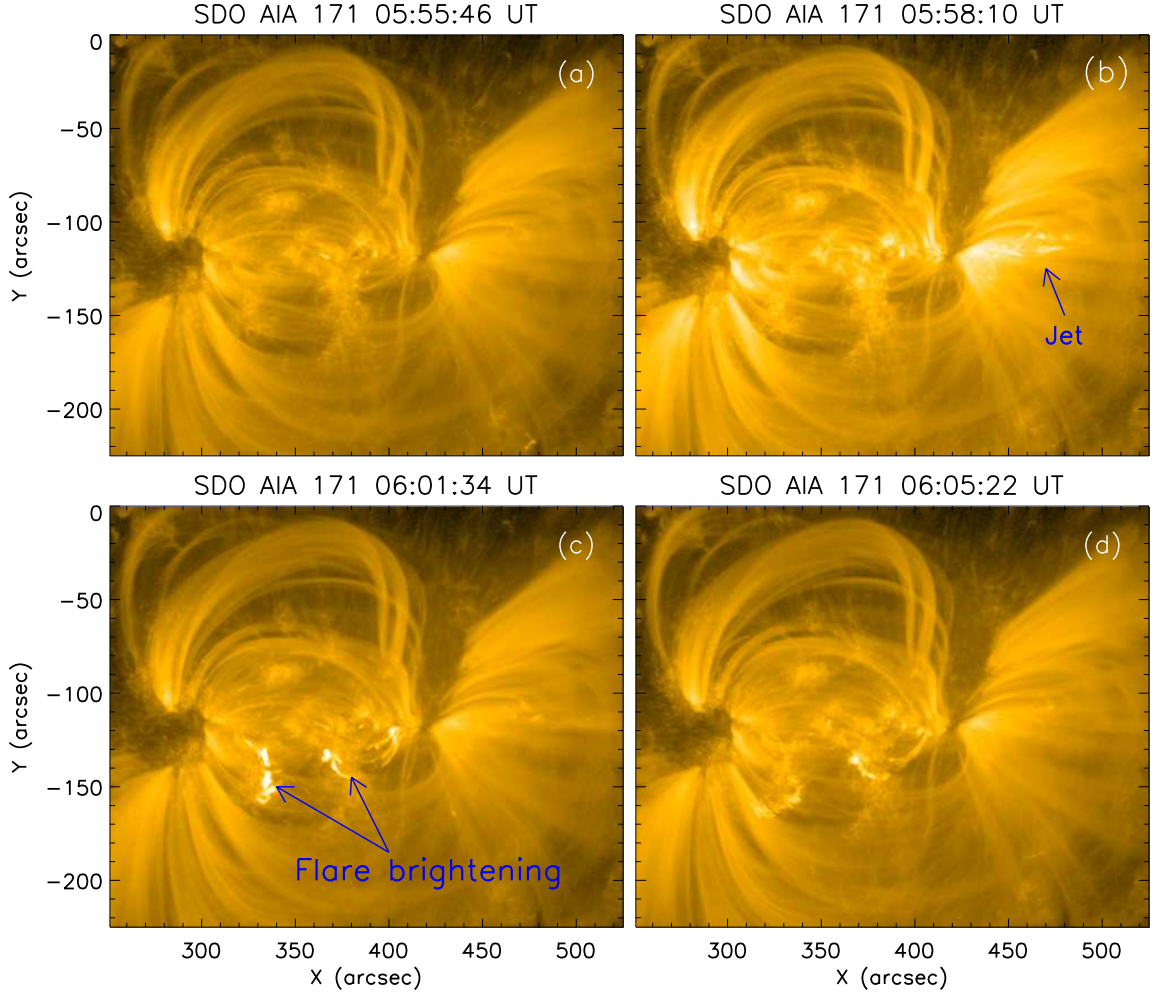}
\caption{A few representative images showing the jet and flare evolution in AIA~171~\AA~images.  
\label{fig:aia171}}
\end{figure}

\begin{figure}[ht!]
\plotone{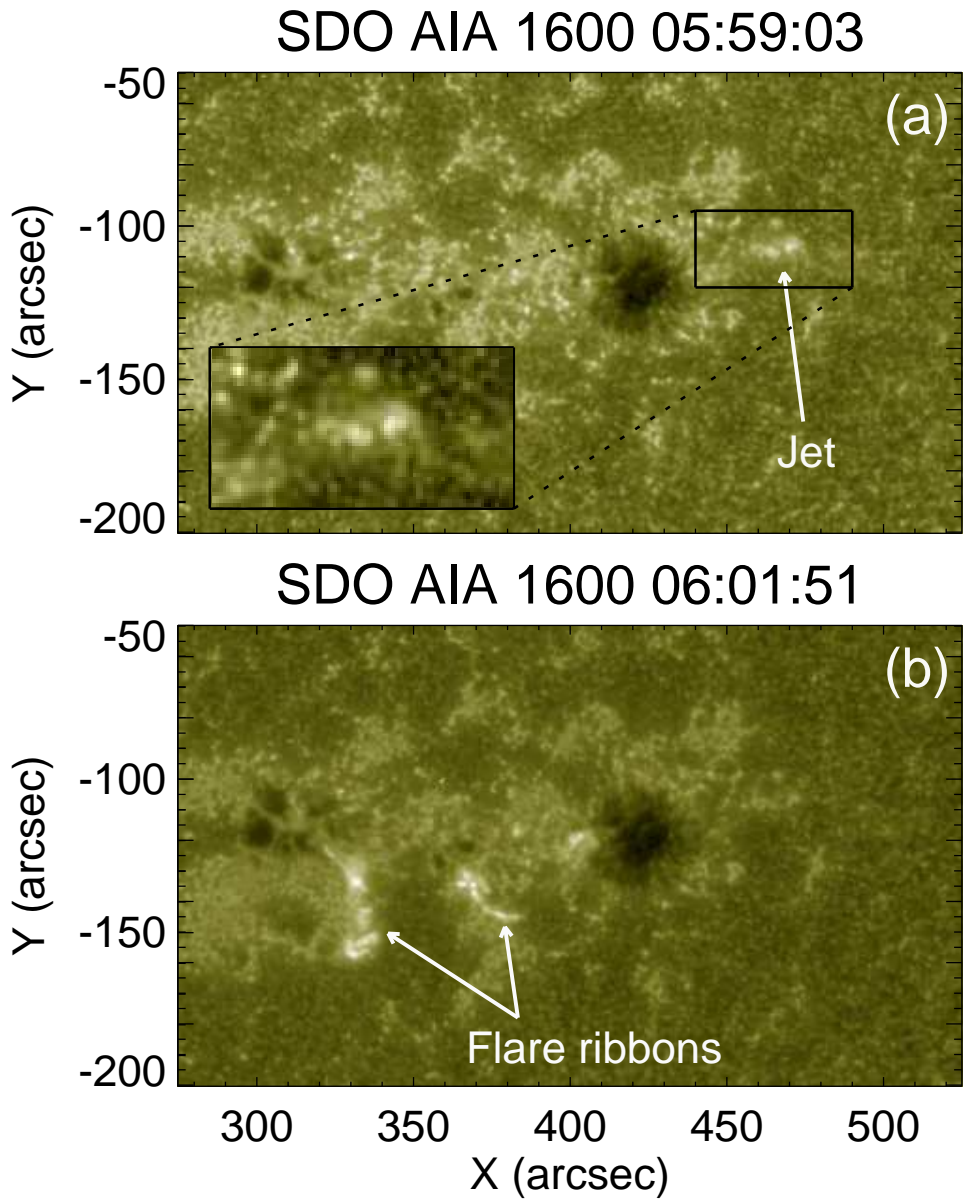}
\vspace{-0.5cm}
\caption{AIA~1600~\AA~images showing the jet activity and subsequent flare developments. It is noted that compared to 304 and 94~\AA\ images, the jet activity in 1600~\AA~images (marked by an arrow in panel (a)) is observed at a later time (only after $\sim$05:58~UT) and also sustained for a brief period. The jet structure at 1600~\AA~is comprised of two kernel-like brightenings. Inlet in panel (a) shows the zoomed image of the jet location to show the jet's kernel with clarity. Panel (b) shows the development of flare ribbons following the jet initiation. 
\label{fig:aia1600}}
\end{figure}

\begin{figure}
\plotone{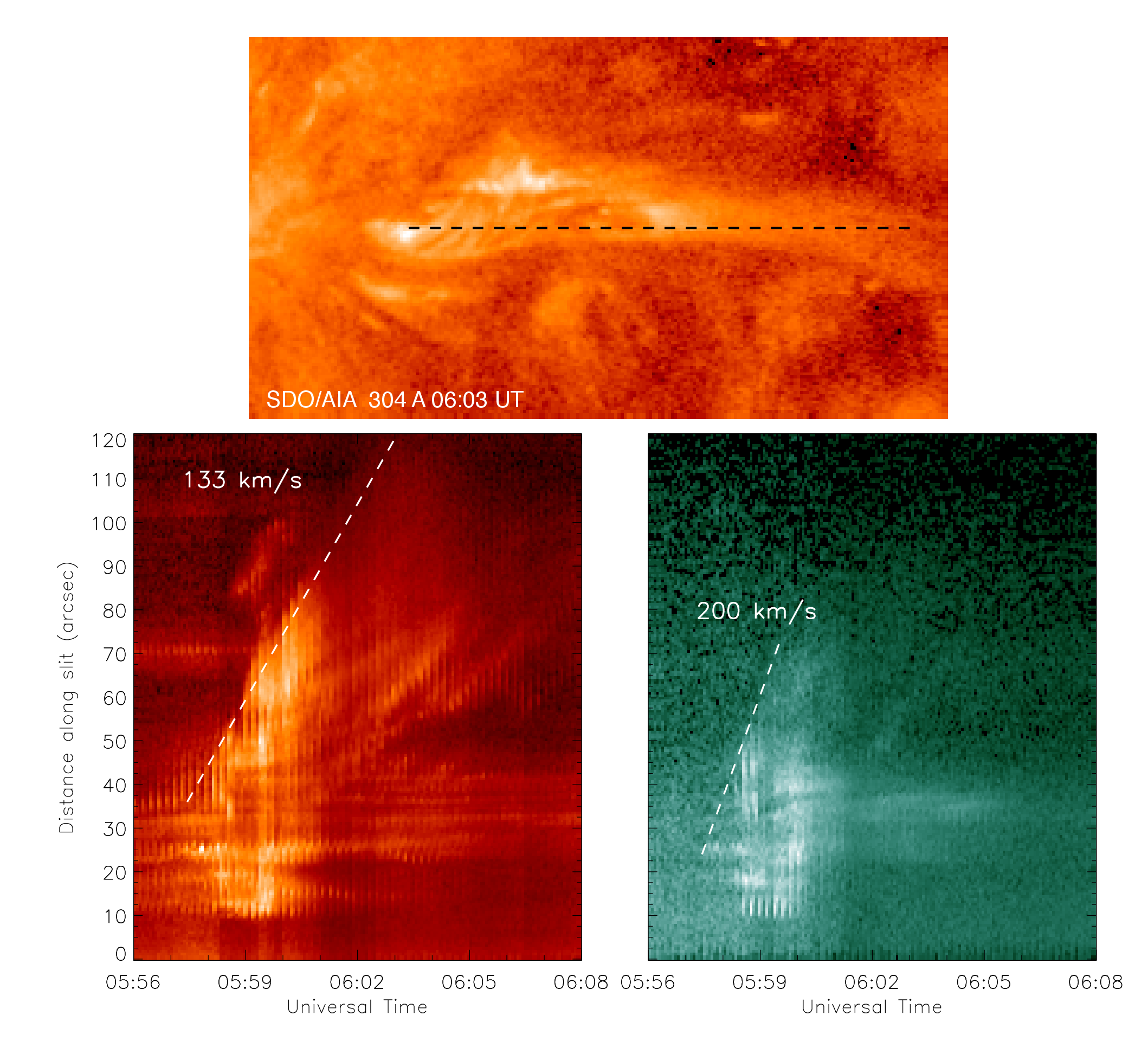}
\caption{Top panel: AIA 304~\AA~image overlaid by the position of an artificial slit indicated by black dashed line. Bottom left and right panels: Time-slice images at 304 and 94~\AA\ respectively showing the evolution of the coronal jet. The average speed of the coronal jet observed at these two wavelengths is annotated in the respective panels.
\label{fig:ala_time_slice}}
\end{figure}

Careful study of the associated coronal EUV emission from the AIA 94~\AA~images (Figure~\ref{fig:aia94}) indicates that the active region corona was quiet until $\sim$05:57~UT, at a time when the jet activity had already begun
We notice jet related developments from $\sim$05:57~UT, onward with the rise of bright, diffuse emission from a localized region (marked by an arrow in Figure~\ref{fig:aia94}(b)). Between 05:58~UT and 06:00~UT, the jet structure brightens up and grows in length while its width remains nearly unchanged. In the subsequent images, we clearly observe eruption of hot plasma along a narrow path with bright and diffuse emission. Similar to 304~\AA~observations, 
we notice a continuous stream of plasma from the base of the jet which resulted in the formation of the jet's spire. At this stage, we observe flare-associated brightenings in the active region (Figure~\ref{fig:aia94}(f)). Notably, the GOES light curves (Figure~\ref{fig:goes_lc}) show a rapid enhancement in the SXR flux only after $\sim$06:00 UT. We further observe fast rise in the intensity of the emission as well as the area of the flaring region (Figure~\ref{fig:aia94}(g)-(i)). The region is enveloped by a system of bright post-flare loops after the flare maximum at $\sim$06:07~UT (Figure~\ref{fig:aia94}(j)). By this time, the jet activity has completely faded away.

In Figure~\ref{fig:aia171}, we show a few representative images taken in the AIA 171~\AA~(Fe~{\footnotesize IX}; log(T) = 5.8) channel. The emission at this
wavelength originates in the quiet corona and upper transition
region. Similar to observations at the 304 and 94~\AA~channels, all the key components of the coronal jet and associated flaring activity, viz., the jet's base and spire along with flare ribbons, are observed in the 171~\AA~images also. Here we emphasize that in 94 and 171~\AA~images, we observe predominantly a hot plasma eruption along the jet's spire while 304~\AA~images show eruption of both hot and cool material distinctly.   

The observations of active region in the AIA~1600~\AA~channel reveal some important features regarding the evolution of the flare and jet emissions at lower heights of the solar atmosphere (Figure~\ref{fig:aia1600}). The AIA~1600~\AA~channel (C~{\footnotesize IV + cont.}; log(T)=5.0) observes combined emission from the transition region and the upper photosphere. In comparison to the 304 and 94~\AA~AIA images, the jet appears to be very faint in 1600~\AA~images and presents two kernel-like structures (Figure~\ref{fig:aia1600}(a)). More importantly, in 1600~\AA~images, the jet is observed about 10 minutes after its first appearance in 304~\AA~observations. Similar to 304~\AA~images, here also we find development of a pair of flare ribbons following the activation of the jet (Figure~\ref{fig:aia1600}(b)). The flare ribbons are morphologically similar in 304 and 1600~\AA~images, although they appear for a shorter time interval at 1600~\AA~images. 

In Figure~\ref{fig:ala_time_slice}, we present time-slice diagrams of the 304 and 94~\AA~images. We notice the jet to be much more prominent at 304~\AA~images over 94~\AA~measurements. From the time-slice plots, we derive a projected speed of 133~km~s$^{-1}$ and 200~km~s$^{-1}$ at chromospheric and coronal temperatures, respectively. Further, we note that the jet structure could be followed in 304~\AA~images up to longer duration and larger projected heights. 

\subsection{Time evolution of flare/jet-associated X-ray and (E)UV emission}
\begin{figure}[ht!]

\plotone{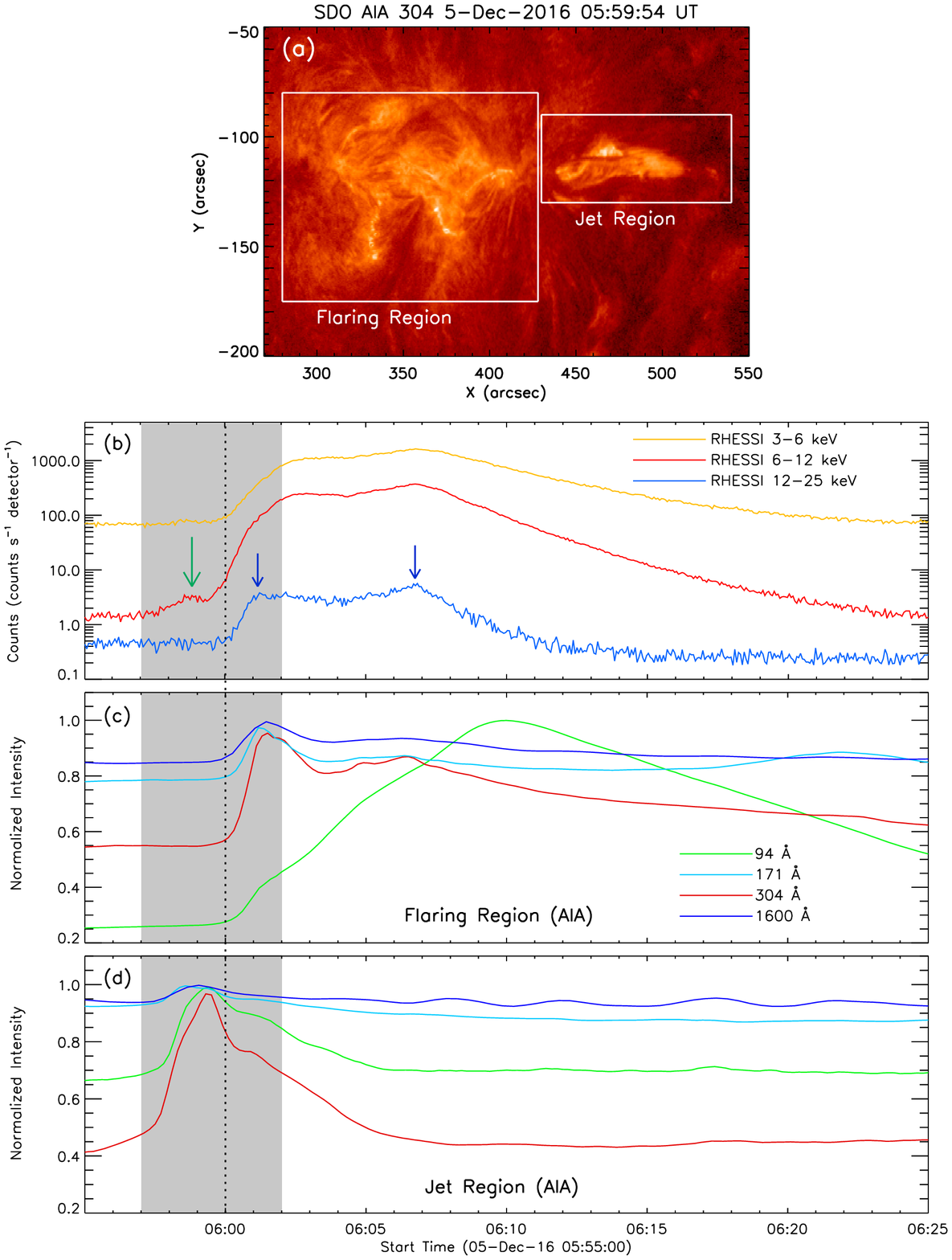}
\caption{Panel (a): AIA 304~\AA~image showing the flaring and jet regions which are selected to compute AIA light curves given in panel (c) and (d). Panel (b): RHESSI X-ray light curves in 3-6, 6-12, and 12-25 keV energy bands. During this observing period, RHESSI's attenuator state was A0, i.e., RHESSI observed solar X-ray emission with its highest sensitivity at low energies. Vertical arrows indicate dual peaks in 12--25~keV light curve at $\sim$06:01 and $\sim$06:07~UT. For clarity of presentation, we have scaled RHESSI count rates by factors of 1, 1/10 and, 1/30 for 3--6, 6--12, and 12--25 keV energy bands, respectively. Panel (c): (E)UV light curves for the flaring region (indicated in AIA~304~\AA~image in panel (a)) in 94, 171, 304, and 1600~\AA\ channels. Panel (d): (E)UV light curves for different channels for the jet region (indicated in AIA~304~\AA~image in panel (a)). (E)UV light curves represent total intensity of the regions of interest, normalized by the peak intensity over the event duration of the respective channels.
The grey shaded region in RHESSI and AIA light curves indicate the period during which the emission from the jet region was pronounced. The dotted line indicate the flare onset time as revealed by GOES 1--8~\AA~soft X-ray measurements. 
\label{fig:rhessi_aia_lc}}
\end{figure}

In Figure~\ref{fig:rhessi_aia_lc}, we present light curves from the entire flaring region (derived from RHESSI and AIA observations) and jet region (derived from AIA observations). The flaring and jet regions, selected for computing E(UV) intensity profiles from AIA data, are shown in Figure~\ref{fig:rhessi_aia_lc}(a).  

Figure~\ref{fig:rhessi_aia_lc}(b) presents RHESSI time profiles in 3--6, 6--12, and 12--25 keV energy bands. In general, all RHESSI light curves clearly indicate a sharp rise in X-ray counts from $\sim$06:00~UT. We note distinct peaks in the time evolution of X-ray flux, a weaker one around 06:01~UT and a stronger second one around 06:07~UT. The dual peak structure becomes more prominent as higher energy bands are considered, i.e., this feature is most pronounced for the 12--25 keV light curve (indicated by blue vertical arrows at $\sim$06:01 and $\sim$06:07~UT in Figure~\ref{fig:rhessi_aia_lc}(b)). 

The (E)UV light curves for the flare and jet region (see Figure~\ref{fig:rhessi_aia_lc}(a)) are shown in panels (c) and (d), respectively, of Figure~\ref{fig:rhessi_aia_lc}. 
The flare-related curves deduced from 171, 304, and 1600~\AA\ observations, show an abrupt increase in intensity with the onset of flare's impulsive phase after $\sim$06:00~UT, similar to RHESSI X-ray light curves.
Furthermore, the increase in the flare emission is similarly steep for the different considered temperature regimes, and corresponding peaks appear nearly co-temporal. In contrast, the flare-region integrated 94~\AA\ emission, shows a clearly different time-dependent pattern. Here, we observe only a gradual, more or less continuous enhancement until $\sim$06:10~UT, clearly representing the heating of the plasma to high temperatures within the flare loops (cf. Figure~\ref{fig:aia94}(h)-(j)). 

From Figure~\ref{fig:rhessi_aia_lc}(d), we clearly notice a significant rise in the emission from the jet region between $\sim$05:57 and $\sim$06:02~UT in all the AIA channels (this interval is shown by the grey shaded area in Figure~\ref{fig:rhessi_aia_lc}(b)-(d)).
Importantly, the jet-related chromospheric emission (304~\AA) starts to increase around 05:54~UT, i.e., with the initial appearance of the early-jet loop like structure (compare Figure~\ref{fig:pre-jet_AIA_HMI}).
Here it is worth to mention a subtle increase in RHESSI 6--12 keV count rates during $\sim$05:57--06:00 UT (marked by a green arrow in Figure~\ref{fig:rhessi_aia_lc}(b)) which likely reflects a response to the jet activity. The flare starts just after the peak phase of the jet emission. In fact, the first peak of flare ($\sim$06:01 UT) is observed when the plasma eruption from the jet's spire was underway.  

\subsection{Spatial evolution of X-ray emission}

\begin{figure}[ht!]
\plotone{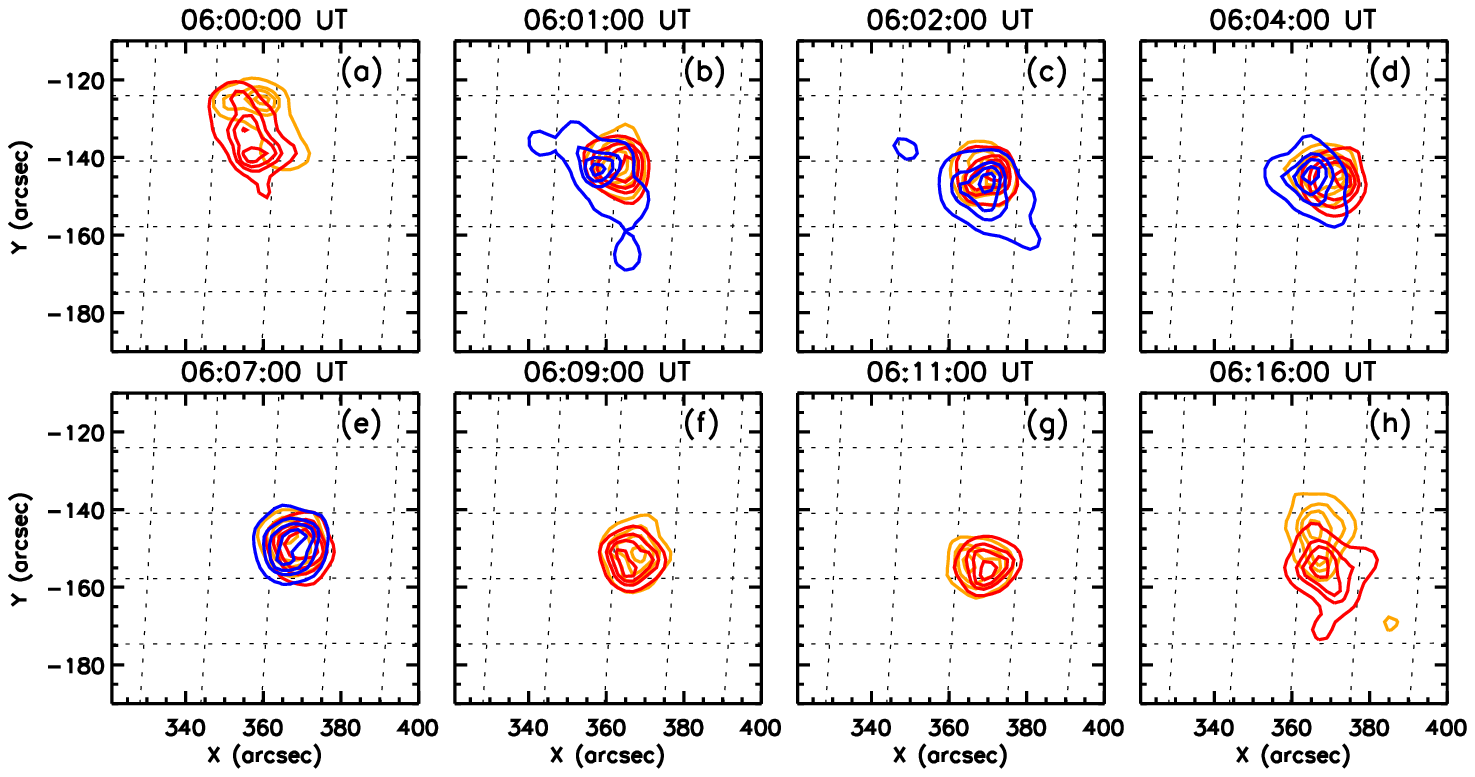}
\caption{Temporal evolution of RHESSI HXR sources in 3--6 (yellow), 6--12 (red), and 12--25 (blue) during different stages of the flare evolution. These images are reconstructed with the CLEAN algorithm. 
The contour levels are set as 55\%, 75\%, 85\%, and 95\% of the peak flux of each images. \label{fig:rhessi_sources}}
\end{figure}

The morphology and evolution of the X-ray sources are presented in Figure~\ref{fig:rhessi_sources}. 
To show the location of the X-ray sources with respect to the spatial attributes of the jet and flare activity, we overplot RHESSI images on selected AIA 304~\AA~(Figure~\ref{fig:aia304}(g)) and 94~\AA~(Figure~\ref{fig:aia94}(g) and (j)) images. We find the appearance of 3--6 and 6--12~keV X-ray sources only after 06:00~UT (Figure~\ref{fig:rhessi_sources}(a)). It is noteworthy that although the RHESSI 6--12 keV light curve shows a slight enhancement in count rates between $\sim$05:57 and 06:00~UT (marked by green arrow in Figure~\ref{fig:rhessi_aia_lc}(b)), the count statistics is probably not sufficient to construct clear source structures during this interval. At 12--25 keV energies, i.e., the highest energy band for this flare, clear sources developed during $\sim$06:01--06:07~UT (Figure~\ref{fig:rhessi_sources}(b)-(e)).
It is interesting that the lower and higher energy X-ray sources are almost co-spatial. Comparison between the location of X-ray emission centroids with the EUV flaring region at 304~\AA~(Figures~\ref{fig:aia304}(g)) clearly indicate that the X-ray emission originates between the two flare ribbons and is slightly more inclined toward the western flare ribbon. The co-spatiality among X-ray emission sources at different energy bands along with their location with respect to the 94~\AA~(Figure~\ref{fig:aia94}(g) and (j)) images reveal that X-ray sources represent emission from hot, low-lying coronal loops that connect the two flare ribbons.

\section{Magnetic field modeling}
\label{sec:mag_model}

\subsection{Jet area}

\begin{figure}[ht!]
\plotone{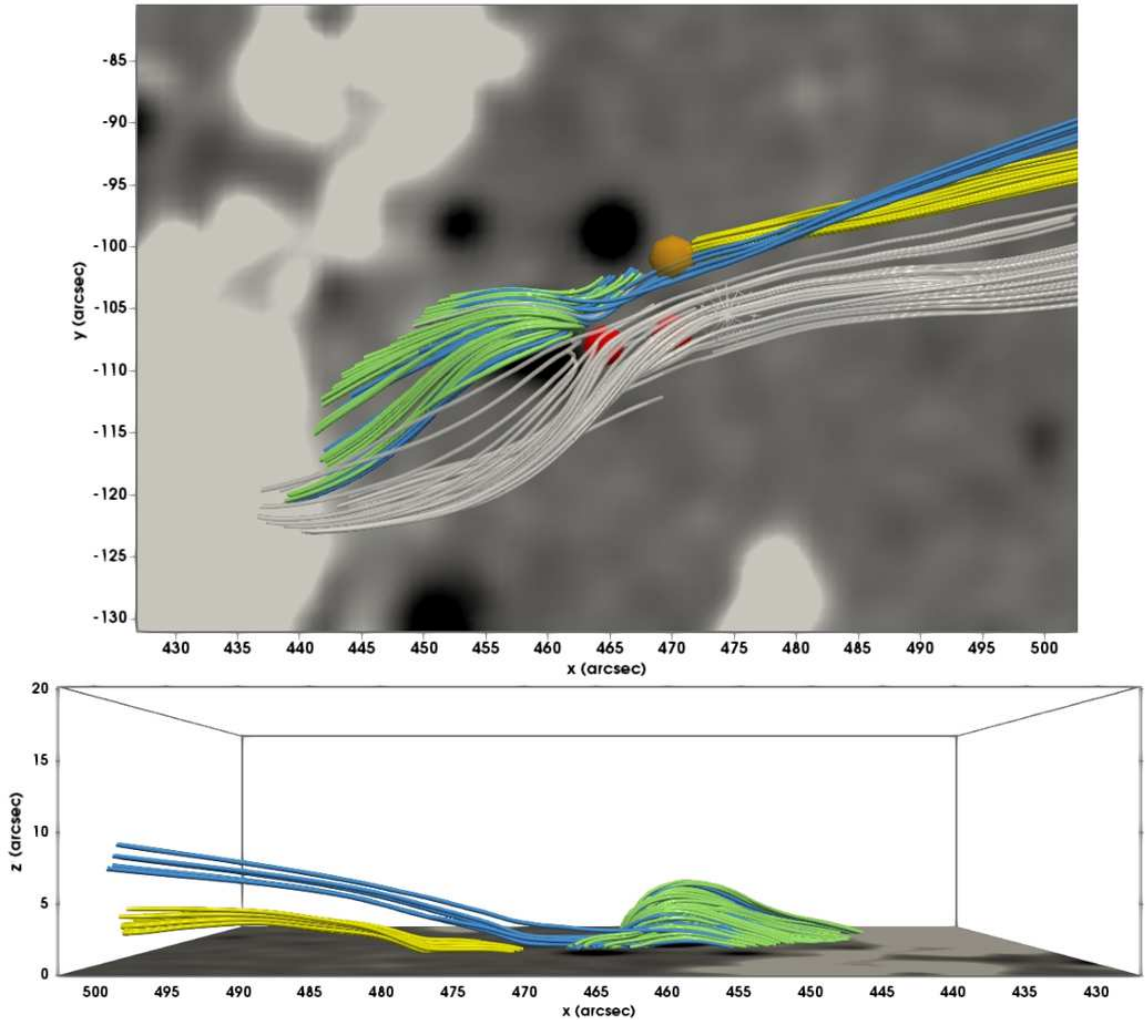}
\caption{Pre-jet (at 05:47~UT; green and yellow lines) and post-jet (at 05:59~UT; blue lines) coronal model magnetic field structure. Grey lines are shown to complete the representation of the jet-related structure as a whole. Grey-scale background resembles the pre-jet vertical magnetic field at the lower model boundary, scaled to $\pm100$~G. The orange colored sphere marks the location of the first compact loop-like brightening observed in AIA 304~\AA, observed around 05:48~UT. The red spheres mark the position of flare kernels observed in AIA 1600~\AA, at 05:59~UT. Units are arcseconds from solar disk center. {\it Bottom panel:} Side view along solar-$y$ direction. Units are arcseconds.
\label{fig:jet_config}}
\end{figure}

As typical for blowout jets, an initial compact brightening in the form of a bright loop-like structure is observed as early as 05:48~UT at the location $(x,y)\approx(468\farcs,-101\farcs)$ (see Figure \ref{fig:pre-jet_AIA_HMI} and Section~\ref{sec:EUV_imaging}; indicated by the orange sphere in upper panel of Figure~\ref{fig:jet_config}). In its vicinity, we find the pre-jet model field terminating in the neighboring negative polarity to the east (green lines) and emerging from the positive polarity to its west (yellow lines). 
Field lines calculated from the same footpoint locations during the post-jet stage (blue lines) suggest a reconnection-based opening of the coronal magnetic field. The lower panel of Figure~\ref{fig:jet_config} shows the jet-associated magnetic field configuration when viewed along the negative solar-$y$ direction. 
In the later stages jet-related bright kernels appear in 1600~\AA\ images (their location marked by red spheres in upper panel of Figure~\ref{fig:jet_config}; also see Figure~\ref{fig:aia1600}(a)). The NLFF model field lines in their vicinity (gray lines) well outline the southern system of bright emission forming the full collimated jet structure (cf. Figure~\ref{fig:rhessi_aia_lc}(a)).

\subsection{Two-ribbon flare area}

\begin{figure}[ht!]
\plotone{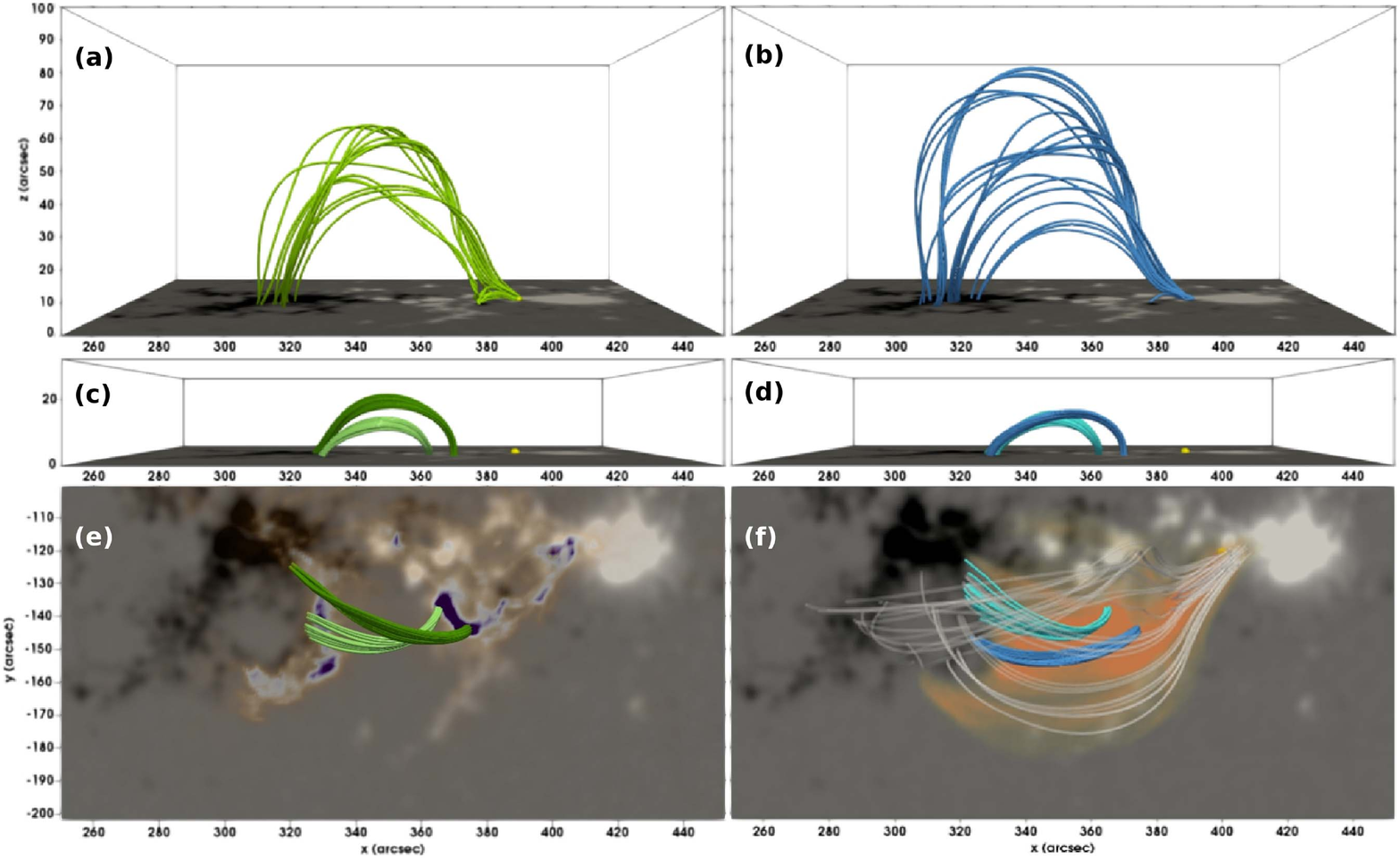}
\caption{Top panels: (a) Pre-flare (at 05:47~UT) and (b) post-flare (at 06:23~UT) coronal model magnetic field structure, originating from the near-spot brightening. The yellow colored sphere marks the location of the inter-spot brightening observed in AIA 304~\AA~at 06:11~UT (see panel (e)). Middle panels: Side view onto the (e) pre- and (f) post-flare coronal field, connecting flare ribbon locations.  Bottom panels: View along vertical direction onto the (e) pre- and (f) post-flare coronal field, connecting flare ribbon locations. Field lines are calculated from two exemplary locations of strongest flare ribbon emission. The yellow and purple contours in (e) outline the flare ribbon emission seen in AIA 304~\AA\ images at 06:11~UT. The orange contours in (f) outline the AIA 94~\AA\ emission observed at 06:11~UT. Grey lines are shown to complete the representation of the flare-related structure and are calculated from the vicinity of the inter-spot brightening. Grey-scale background resembles the pre- and post-flare vertical magnetic field at the lower model boundary, scaled to $\pm500$~G. Units are arcseconds.
\label{fig:flare_config}}
\end{figure}

The first notable flare-related brightening is observed at around 05:50~UT in AIA 304~\AA\ images (i.e., about 2 minutes after the first signatures of the jet). It is located at $(x,y)\approx(400\farcs,-121\farcs)$, just east of the positive-polarity sunspot (indicated by a yellow sphere in Fig.~\ref{fig:flare_config}(a), (c), (d); see the location marked in Figure~\ref{fig:aia304}(a) as early flare brightening). After that, a number of flows are observed, towards the south-east, where the flare ribbons are formed at later stages. A system of quasi-parallel ribbons develops from $\sim$05:59~UT onward (Figure~\ref{fig:aia304}(e)-(h)).

The pre- and post-flare coronal magnetic field, rooted in the near-spot brightening (yellow sphere in Figure~\ref{fig:flare_config}(a), (c), (d)) is shown in Figure~\ref{fig:flare_config}(a) and \ref{fig:flare_config}(b), respectively. The previously closed magnetic field rooted in the small negative-polarity region located around $(x,y)=(385\farcs,-135\farcs)$, appears to have been opened up in the course of the flare, possibly allowing the underlying core-field to expand to greater heights. The model field originating from the vicinity of the near-spot brightening (semi-transparent lines in Figure~\ref{fig:flare_config}(f)) well outline the hot flare  emission observed in AIA 94~\AA\ (filled contours in Figure~\ref{fig:flare_config}(f)), and nicely connects it to beyond the site where the two-ribbon flare developed.

A side view on the pre- and post-flare model fields rooted in flare kernels/ribbons (as observed at 06:11~UT and shown as colored contours in Figure~\ref{fig:flare_config}(e)) are shown in Figure~\ref{fig:flare_config}(c) and \ref{fig:flare_config}(d), respectively. Since traced from locations populated by flare kernels/ribbons, they represent magnetic field structures that have been possibly involved in magnetic reconnection. Traced from the same footpoint locations, the pre-flare configuration exhibits an ``X-shape'', while the post-flare configuration appears more or less parallel oriented. The X-shape is also visible in the flare emission observed at 06:04~UT (Figure~\ref{fig:aia94}(i)).

\section{Discussions}
\label{sec:discussion}

\begin{table}
\caption{Chronology of events during the coronal jet and subsequent C1.2 flare on 2016 December 5}
\begin{tabular}{p{1.5in}p{0.5in}p{0.8in}p{3.8in}}
\tableline\tableline
Observations  & Time (UT) & Observing wavelength & Remarks\\
\tableline

Jet initiation & 05:48 & AIA 304~\AA & Compact loop-like brightening is observed which later developed to jet's base\\

Early flare brightening & 05:50 & AIA 304~\AA & Occurrence of multiple, kernel-like brightenings in the flaring region; several flows are observed where flare ribbons subsequently formed \\  

Rapid expansion of jet's base and onset of eruption & 05:54 & AIA 304~\AA& Collimated eruption of hot and cool plasma until $\sim$06:15~UT; Clear jet structure with distinct appearance of its base and spire\\

Development of parallel flare ribbons & 05:59 & AIA 304~\AA & Separation between ribbons remains nearly unchanged\\

Development of hot post-flare loops and beginning of X-ray emission & 06:00 & AIA 94~\AA, RHESSI~3--12 keV& GOES observations show flare from $\sim$06:00--06:20~UT; HXR emission up to 25 keV; Single, compact HXR sources that remain cospatial to hot EUV loops\\  
\tableline
\label{tab:summary}
\end{tabular}
\end{table}

In this paper, we present a multi-wavelength investigation of a very intriguing coronal jet in AR 12615 that is followed by a C1.2 two-ribbon flare on 2016 December 5. The study explores the fundamental processes associated with the causes and consequences of the jet activity. In Table~\ref{tab:summary}, we give a chronological list of events and summarized the prominent multi-wavelength observations during various stages.

The jet initiation is first observed in AIA~304~\AA~observations at $\sim$05:48~UT in the form of a compact loop-like brightening. We observe a rapid enhancement in the intensity and area of the region after $\sim$05:54~UT as it undergoes a fast lengthening over lateral expansion (Figure~\ref{fig:aia304}). Within a couple of minutes, a clear typical jet structure emerges with collimated eruption of plasma from jet's base which forms its spire. From AIA~304~\AA~images, we note the spire to be narrow in the beginning but then broadens out with time. 
Based on the morphological characteristics along with the dynamical evolution of the jet, we identify this to be a blowout jet \citep{Moore2010,Moore2013}.
The jet activity is also seen at other coronal temperatures, although with some differences in timing and spatial organization, providing evidence for the erupting plasma to be multi-thermal. Notably, in hot coronal 94~\AA\ emission (characteristic temperature: 6.3 $\times$ 10$^{6}$~K) the jet onset appears about 3 minutes delayed with respect to the initially observed jet activation at chromospheric temperatures (as observed in 304~\AA\ images at a characteristic temperature of 50,000~K). The different appearance of the jet in these two characteristic temperature regimes provides important insights regarding the underlying physical processes.
Apparently, the early reconnection causing the jet's onset occurred at chromosphere/transition-region heights rather than in the corona. The spatial and temporal changes during the early phase of the jet activity at 
chromospheric temperatures 
show that the reconnections were very localized and of smaller-scale.
Only after a considerable increase of the jet base area (and associated increase in chromospheric emission), collimated eruptive hot plasma becomes visible also at coronal temperatures (i.e., at 94~\AA; cf. Figures~\ref{fig:aia304} and \ref{fig:aia94}). 
At this time the spire broadens out and both cool and hot components of erupting plasma are observed. These observations suggest that now the magnetic reconnections are no longer restricted to the localized region of chromosphere but occur simultaneously at multiple locations in the chromosphere and upward (i.e., transition region and corona).     

From AIA 304~\AA~ images, we note an elongated, cool plasma structure, lying along the jet (see jet activation region in Figure~\ref{fig:aia304}(c)). Interestingly, a blob of cool plasma erupted from this region (indicated by arrows in the inlet of Figure~\ref{fig:aia304}(d)) which is followed by lengthening and broadening of the jet's spire (Figure~\ref{fig:aia304}(d)-(e)). Complementary to the present observations, we note earlier studies that reveal eruption of cool plasma material as mini-filament structures, in conjunction to the EUV/X-ray jets \citep{Sterling2015,Sterling2016}.
We note that as the ejection of cooler plasma
proceeded westward, bright emission emerges from its front portion along with the continuation of early brightenings at the base of the jet (regions of enhanced emission at the base and front of the jet are indicated as R1 and R2, respectively, in Figure~\ref{fig:aia304}(e)). We interpret the bright emission from the R2 as a consequence of magnetic reconnection triggered by the mini-filament eruption. It is likely that latter widespread reconnections heat the cool plasma to million degrees which eventually gives rise to the jet activity observed at the hot 94~\AA\ EUV channel. In view of this, we propose that the early brightening forming the jet base is the result of spontaneous reconnection between the pre-existing large-scale magnetic structure and a small bipolar magnetic region, as proposed in the standard jet model \citep[see e.g.,][]{Shimojo2000,Canfield1996}. This would not only provide initial opening for the outward flow of hot and cool plasma but also weaken the inner core fields of the active region. Such a scenario is supported by the NLFF magnetic field model of the jet area during the pre-jet phase that show a favorable magnetic configuration (Figure~\ref{fig:jet_config}) in which a system of closed field line associated with a photospheric bipolar region (green lines) existed in the vicinity of an open field structure (yellow lines). Change in magnetic field configuration following the jet (blue lines) implies 
opening of previously closed coronal magnetic fields.
The NLFF model results suggest a rather inclined/low-lying jet body which is consistent with the AIA observations.

The temporal correlation between the jet and flare activity is clearly seen from the X-ray and (E)UV light curves (Figure~\ref{fig:rhessi_aia_lc}). The flare initiates and two parallel flare ribbons grow as the collimated eruption proceeds through the jet's spire (see Figure~\ref{fig:aia304}(e)-(h)). 
Notably, intensity of flare ribbons increases without any noticeable change in their spatial separation, typical for confined flares \citep{Kushwaha2014}. The X-ray light curves exhibit a dual-peak profile.
The first peak shows an energy dependent structure, i.e., the emission in 12-25 keV peaks earlier than in the low-energy bands. The second X-ray peak is observed nearly simultaneously in all the energy bands. The structure and morphology of flare ribbons and post-flare loops (cf. Figure~\ref{fig:AR_mag_euv}(c)-(d)) suggest that the flare is caused due to magnetic reconnection involving the magnetic loops that connect the trailing sunspot group of negative polarity (N) with the two well-separated leading sunspot groups of positive polarities (P1 and P2) (see Figure~\ref{fig:AR_mag_euv}(d)). 
Although association of jet activity with compact flare brightening is not uncommon \citep{Shimojo2000,Chandra2017}, the present study demonstrate the association of a blowout jet with a well developed two-ribbon flare. 
Ribbons are considered as the most prominent signatures of the standard flare as they correspond to the chromospheric footpoints of coronal loops that have reconnected \citep{Lin2003}. 

The standard flare scenario during the C1.2 flare is manifested not only in terms of the development of the flare ribbons but also in the formation of a bright post-flare loop system that eventually envelops the whole flaring region (Figure~\ref{fig:aia94}(i)-(j)). As discussed earlier, RHESSI light curves (also see GOES profiles) exhibit two peaks at an interval of about $\sim$5~minutes (Figures~\ref{fig:goes_lc} and \ref{fig:rhessi_aia_lc}(b)) with the latter peak being the overall X-ray flare maximum. The multi-channel X-ray images reveal X-ray emission at different energy bands (3--6, 6--12, and 12--25 keV) to be cospatial. Further, the HXR sources are spatially associated with the brightest part of the EUV post-flare loop system. Therefore, it is reasonable to assume that almost all X-ray emission during the flare originated from the newly formed loop system, i.e., the contribution of X-rays from footpoint regions (which are delineated by the flare ribbons) is observationally insignificant. Insignificant HXR footpoint emission could be attributed to weak particle precipitation along the closed post-flare loops or higher density in the overlying coronal loop system \citep{Veronig2004,Fletcher2011}. The higher coronal densities would offer greater efficiency to collisionally stop most of the accelerated electrons before they reach the lower chromospheric and transition region heights \citep{Veronig2004}. By combining the temporal and spatial characteristics of flare emission from RHESSI and AIA, we conclude that the first peak essentially represent the flare reconnection, possibly triggered by the jet activity, while the second peak is attributed to intense thermal emission from closed, post-reconnected flare loops.

The coronal magnetic field rooted in flare ribbons, derived from a NLFF model, suggests the pre-flare configuration to exhibit an ``X-shape" which is susceptible to magnetic reconnection (left panels of Figure~\ref{fig:flare_config}). On the other hand, the post-flare configuration appears more or less parallel oriented as expected following the magnetic reconnection (right panels of Figure~\ref{fig:flare_config}). Notably, the AIA 94~\AA~images during early and post-flare phases are in good agreement with the model field configuration.   
Possible reasons for the initiation of the reconnection might include the elevation and consequent interaction of the X-shaped flux system, due to a change in the surrounding (and/or overlying) field. Since we were not able to deduce a direct magnetic link between the jet area and the two-ribbon flare region, the possible influence of the former on the latter might be an indirect one. It is very likely that the jet activity may have driven the meta-stable pre-flare magnetic flux system further away from an equilibrium state, due to a disturbance to the overall AR coronal magnetic field configuration, thus indirectly inducing the flare activity on the other side of the sunspot. 
Direct evidence, however, remains elusive.

\section{Conclusions}
\label{sec:conclusion}
Major highlights of our investigation are as follows:

\begin{itemize}
\item We have studied the multi-wavelength evolution of a coronal jet which turns out to be of blowout category. The initial compact brightening spreads in a larger region and collimated eruption of hot and cool plasma is observed. 

\item The NLFF modeling of coronal magnetic fields suggest magnetic reconnection opening of closed, low-lying bipolar fields with nearby open flux system. The model suggests a rather inclined/low lying jet body which nicely matches with the observations.

\item A C1.2 flare initiated just after the jet onset. The standard two-ribbon structure developed when the eruption of the multi-thermal plasma was underway through the jet's spire. The temporal correlations and spatial correspondence   between the jet and flare activities point toward a possible association between the two phenomena.

\item
The coronal magnetic field structure extrapolated by NLFF model successfully demonstrate the pre-flare phase to exhibit an ``X-type" configuration while the fields lines appear to me more or less parallel oriented following the flare.   

\item
By combining the results of observations and modeling, we propose that the jet has likely triggered the C1.2 flare by perturbing the meta-stable ``X-type" pre-flare magnetic flux system further away from the equilibrium.    

\end{itemize}

\acknowledgments
We thank the SDO and RHESSI teams for their open data policy. SDO and RHESSI are NASA's missions under Living With a Star (LWS) and SMall EXplorer (SMEX) programs, respectively. This work is supported by the Indo-Austrian joint research project no. INT/AUSTRIA/BMWF/P-05/2017 and OeAD project no. IN 03/2017. A.M.V. and J.K.T. acknowledge the Austrian Science Fund (FWF): P27292-N20. We
thank the anonymous referee for providing useful comments and suggestions.

\newpage
\bibliographystyle{apj}
\bibliography{New}

\begin{thebibliography}{}
\expandafter\ifx\csname natexlab\endcsname\relax\def\natexlab#1{#1}\fi

\bibitem[{{Alexander} \& {Fletcher}(1999)}]{Alexander1999}
{Alexander}, D., \& {Fletcher}, L. 1999, \solphys, 190, 167

\bibitem[{{Benz}(2017)}]{Benz2017}
{Benz}, A.~O. 2017, Living Reviews in Solar Physics, 14, 2

\bibitem[{{Canfield} {et~al.}(1996){Canfield}, {Reardon}, {Leka}, {Shibata},
  {Yokoyama}, \& {Shimojo}}]{Canfield1996}
{Canfield}, R.~C., {Reardon}, K.~P., {Leka}, K.~D., {et~al.} 1996, \apj, 464,
  1016

\bibitem[{{Chandra} {et~al.}(2015){Chandra}, {Gupta}, {Mulay}, \&
  {Tripathi}}]{Chandra2015}
{Chandra}, R., {Gupta}, G.~R., {Mulay}, S., \& {Tripathi}, D. 2015, \mnras,
  446, 3741

\bibitem[{{Chandra} {et~al.}(2017){Chandra}, {Mandrini}, {Schmieder}, {Joshi},
  {Cristiani}, {Cremades}, {Pariat}, {Nuevo}, {Srivastava}, \&
  {Uddin}}]{Chandra2017}
{Chandra}, R., {Mandrini}, C.~H., {Schmieder}, B., {et~al.} 2017, \aap, 598,
  A41

\bibitem[{{DeRosa} {et~al.}(2015){DeRosa}, {Wheatland}, {Leka}, {Barnes},
  {Amari}, {Canou}, {Gilchrist}, {Thalmann}, {Valori}, {Wiegelmann},
  {Schrijver}, {Malanushenko}, {Sun}, \& {R{\'e}gnier}}]{2015ApJ...811..107D}
{DeRosa}, M.~L., {Wheatland}, M.~S., {Leka}, K.~D., {et~al.} 2015, \apj, 811,
  107

\bibitem[{{Fletcher} {et~al.}(2011){Fletcher}, {Dennis}, {Hudson}, {Krucker},
  {Phillips}, {Veronig}, {Battaglia}, {Bone}, {Caspi}, {Chen}, {Gallagher},
  {Grigis}, {Ji}, {Liu}, {Milligan}, \& {Temmer}}]{Fletcher2011}
{Fletcher}, L., {Dennis}, B.~R., {Hudson}, H.~S., {et~al.} 2011, \ssr, 159, 19

\bibitem[{{Gary} \& {Hagyard}(1990)}]{1990SoPh..126...21G}
{Gary}, G.~A., \& {Hagyard}, M.~J. 1990, \solphys, 126, 21

\bibitem[{{Hoeksema} {et~al.}(2014){Hoeksema}, {Liu}, {Hayashi}, {Sun},
  {Schou}, {Couvidat}, {Norton}, {Bobra}, {Centeno}, {Leka}, {Barnes}, \&
  {Turmon}}]{2014SoPh..289.3483H}
{Hoeksema}, J.~T., {Liu}, Y., {Hayashi}, K., {et~al.} 2014, \solphys, 289, 3483

\bibitem[{{Jiang} {et~al.}(2007){Jiang}, {Chen}, {Li}, {Shen}, \&
  {Yang}}]{Jiang2007}
{Jiang}, Y.~C., {Chen}, H.~D., {Li}, K.~J., {Shen}, Y.~D., \& {Yang}, L.~H.
  2007, \aap, 469, 331

\bibitem[{{Joshi} {et~al.}(2016){Joshi}, {Kushwaha}, {Veronig}, \&
  {Cho}}]{Joshi2016}
{Joshi}, B., {Kushwaha}, U., {Veronig}, A.~M., \& {Cho}, K.-S. 2016, \apj, 832,
  130

\bibitem[{{Joshi} {et~al.}(2017){Joshi}, {Schmieder}, {Chandra}, {Aulanier},
  {Zuccarello}, \& {Uddin}}]{JoshiR2017}
{Joshi}, R., {Schmieder}, B., {Chandra}, R., {et~al.} 2017, ArXiv e-prints,
  arXiv:1709.02791

\bibitem[{{Kushwaha} {et~al.}(2014){Kushwaha}, {Joshi}, {Cho}, {Veronig},
  {Tiwari}, \& {Mathew}}]{Kushwaha2014}
{Kushwaha}, U., {Joshi}, B., {Cho}, K.-S., {et~al.} 2014, \apj, 791, 23

\bibitem[{{Lemen} {et~al.}(2012){Lemen}, {Title}, {Akin}, {Boerner}, {Chou},
  {Drake}, {Duncan}, {Edwards}, {Friedlaender}, {Heyman}, {Hurlburt}, {Katz},
  {Kushner}, {Levay}, {Lindgren}, {Mathur}, {McFeaters}, {Mitchell}, {Rehse},
  {Schrijver}, {Springer}, {Stern}, {Tarbell}, {Wuelser}, {Wolfson}, {Yanari},
  {Bookbinder}, {Cheimets}, {Caldwell}, {Deluca}, {Gates}, {Golub}, {Park},
  {Podgorski}, {Bush}, {Scherrer}, {Gummin}, {Smith}, {Auker}, {Jerram},
  {Pool}, {Soufli}, {Windt}, {Beardsley}, {Clapp}, {Lang}, \&
  {Waltham}}]{Lemen2012}
{Lemen}, J.~R., {Title}, A.~M., {Akin}, D.~J., {et~al.} 2012, \solphys, 275, 17

\bibitem[{{Lin} {et~al.}(2003){Lin}, {Soon}, \& {Baliunas}}]{Lin2003}
{Lin}, J., {Soon}, W., \& {Baliunas}, S.~L. 2003, \nar, 47, 53

\bibitem[{{Lin} {et~al.}(2002){Lin}, {Dennis}, {Hurford}, {Smith}, {Zehnder},
  {Harvey}, {Curtis}, {Pankow}, {Turin}, {Bester}, {Csillaghy}, {Lewis},
  {Madden}, {van Beek}, {Appleby}, {Raudorf}, {McTiernan}, {Ramaty}, {Schmahl},
  {Schwartz}, {Krucker}, {Abiad}, {Quinn}, {Berg}, {Hashii}, {Sterling},
  {Jackson}, {Pratt}, {Campbell}, {Malone}, {Landis}, {Barrington-Leigh},
  {Slassi-Sennou}, {Cork}, {Clark}, {Amato}, {Orwig}, {Boyle}, {Banks},
  {Shirey}, {Tolbert}, {Zarro}, {Snow}, {Thomsen}, {Henneck}, {McHedlishvili},
  {Ming}, {Fivian}, {Jordan}, {Wanner}, {Crubb}, {Preble}, {Matranga}, {Benz},
  {Hudson}, {Canfield}, {Holman}, {Crannell}, {Kosugi}, {Emslie}, {Vilmer},
  {Brown}, {Johns-Krull}, {Aschwanden}, {Metcalf}, \& {Conway}}]{LinRP2002}
{Lin}, R.~P., {Dennis}, B.~R., {Hurford}, G.~J., {et~al.} 2002, \solphys, 210,
  3

\bibitem[{{Moore} {et~al.}(2010){Moore}, {Cirtain}, {Sterling}, \&
  {Falconer}}]{Moore2010}
{Moore}, R.~L., {Cirtain}, J.~W., {Sterling}, A.~C., \& {Falconer}, D.~A. 2010,
  \apj, 720, 757

\bibitem[{{Moore} {et~al.}(2013){Moore}, {Sterling}, {Falconer}, \&
  {Robe}}]{Moore2013}
{Moore}, R.~L., {Sterling}, A.~C., {Falconer}, D.~A., \& {Robe}, D. 2013, \apj,
  769, 134

\bibitem[{{Nistic{\`o}} {et~al.}(2009){Nistic{\`o}}, {Bothmer}, {Patsourakos},
  \& {Zimbardo}}]{Nistico2009}
{Nistic{\`o}}, G., {Bothmer}, V., {Patsourakos}, S., \& {Zimbardo}, G. 2009,
  \solphys, 259, 87

\bibitem[{{Nistic{\`o}} {et~al.}(2010){Nistic{\`o}}, {Bothmer}, {Patsourakos},
  \& {Zimbardo}}]{Nistico2010}
---. 2010, Annales Geophysicae, 28, 687

\bibitem[{{Patsourakos} {et~al.}(2008){Patsourakos}, {Pariat}, {Vourlidas},
  {Antiochos}, \& {Wuelser}}]{Patsourakos2008}
{Patsourakos}, S., {Pariat}, E., {Vourlidas}, A., {Antiochos}, S.~K., \&
  {Wuelser}, J.~P. 2008, \apjl, 680, L73

\bibitem[{{Pesnell} {et~al.}(2012){Pesnell}, {Thompson}, \&
  {Chamberlin}}]{Pesnell2012}
{Pesnell}, W.~D., {Thompson}, B.~J., \& {Chamberlin}, P.~C. 2012, \solphys,
  275, 3

\bibitem[{{Raouafi} {et~al.}(2016){Raouafi}, {Patsourakos}, {Pariat}, {Young},
  {Sterling}, {Savcheva}, {Shimojo}, {Moreno-Insertis}, {DeVore}, {Archontis},
  {T{\"o}r{\"o}k}, {Mason}, {Curdt}, {Meyer}, {Dalmasse}, \&
  {Matsui}}]{Raouafi2016}
{Raouafi}, N.~E., {Patsourakos}, S., {Pariat}, E., {et~al.} 2016, \ssr, 201, 1

\bibitem[{{Schou} {et~al.}(2012){Schou}, {Scherrer}, {Bush}, {Wachter},
  {Couvidat}, {Rabello-Soares}, {Bogart}, {Hoeksema}, {Liu}, {Duvall}, {Akin},
  {Allard}, {Miles}, {Rairden}, {Shine}, {Tarbell}, {Title}, {Wolfson},
  {Elmore}, {Norton}, \& {Tomczyk}}]{Schou2012}
{Schou}, J., {Scherrer}, P.~H., {Bush}, R.~I., {et~al.} 2012, \solphys, 275,
  229

\bibitem[{{Shibata}(1999)}]{Shibata1999}
{Shibata}, K. 1999, \apss, 264, 129

\bibitem[{{Shibata} {et~al.}(1992){Shibata}, {Ishido}, {Acton}, {Strong},
  {Hirayama}, {Uchida}, {McAllister}, {Matsumoto}, {Tsuneta}, {Shimizu},
  {Hara}, {Sakurai}, {Ichimoto}, {Nishino}, \& {Ogawara}}]{Shibata1992}
{Shibata}, K., {Ishido}, Y., {Acton}, L.~W., {et~al.} 1992, \pasj, 44, L173

\bibitem[{{Shimojo} {et~al.}(1996){Shimojo}, {Hashimoto}, {Shibata},
  {Hirayama}, {Hudson}, \& {Acton}}]{Shimojo1996}
{Shimojo}, M., {Hashimoto}, S., {Shibata}, K., {et~al.} 1996, \pasj, 48, 123

\bibitem[{{Shimojo} \& {Shibata}(2000)}]{Shimojo2000}
{Shimojo}, M., \& {Shibata}, K. 2000, \apj, 542, 1100

\bibitem[{{Shimojo} {et~al.}(1998){Shimojo}, {Shibata}, \&
  {Harvey}}]{Shimojo1998}
{Shimojo}, M., {Shibata}, K., \& {Harvey}, K.~L. 1998, \solphys, 178, 379

\bibitem[{{Sterling} {et~al.}(2015){Sterling}, {Moore}, {Falconer}, \&
  {Adams}}]{Sterling2015}
{Sterling}, A.~C., {Moore}, R.~L., {Falconer}, D.~A., \& {Adams}, M. 2015,
  \nat, 523, 437

\bibitem[{{Sterling} {et~al.}(2016){Sterling}, {Moore}, {Falconer}, {Panesar},
  {Akiyama}, {Yashiro}, \& {Gopalswamy}}]{Sterling2016}
{Sterling}, A.~C., {Moore}, R.~L., {Falconer}, D.~A., {et~al.} 2016, \apj, 821,
  100

\bibitem[{{Tsuneta} {et~al.}(1991){Tsuneta}, {Acton}, {Bruner}, {Lemen},
  {Brown}, {Caravalho}, {Catura}, {Freeland}, {Jurcevich}, {Morrison},
  {Ogawara}, {Hirayama}, \& {Owens}}]{Tsuenta1991}
{Tsuneta}, S., {Acton}, L., {Bruner}, M., {et~al.} 1991, \solphys, 136, 37

\bibitem[{{Veronig} \& {Brown}(2004)}]{Veronig2004}
{Veronig}, A.~M., \& {Brown}, J.~C. 2004, \apjl, 603, L117

\bibitem[{{Wang} \& {Sheeley}(2002)}]{Wang2002}
{Wang}, Y.-M., \& {Sheeley}, Jr., N.~R. 2002, \apj, 575, 542

\bibitem[{{Wang} {et~al.}(1998){Wang}, {Sheeley}, {Socker}, {Howard},
  {Brueckner}, {Michels}, {Moses}, {St.~Cyr}, {Llebaria}, \&
  {Delaboudini{\`e}re}}]{Wang1998}
{Wang}, Y.-M., {Sheeley}, Jr., N.~R., {Socker}, D.~G., {et~al.} 1998, \apj,
  508, 899

\bibitem[{{Wiegelmann} \& {Inhester}(2010)}]{2010A&A...516A.107W}
{Wiegelmann}, T., \& {Inhester}, B. 2010, \aap, 516, A107

\bibitem[{{Wiegelmann} {et~al.}(2014){Wiegelmann}, {Thalmann}, \&
  {Solanki}}]{Wiegelmann2014}
{Wiegelmann}, T., {Thalmann}, J.~K., \& {Solanki}, S.~K. 2014, \aapr, 22, 78

\bibitem[{{Yokoyama} \& {Shibata}(1995)}]{Yokoyama1995}
{Yokoyama}, T., \& {Shibata}, K. 1995, \nat, 375, 42

\end{thebibliography}



\end{document}